\documentstyle[twocolumn,aps,prl,floats]{revtex}
\begin{document}               
\title{Non-Abelian Effects in Gravitation}
\author{A.~Deur}
\address{University of Virginia, Charlottesville, Virginia 22904}
\break \break
\maketitle
\begin{abstract}                

Abstract: \emph{The non-abelian symmetry of a lagrangian invalidates the principle
of superposition for the field described by this lagrangian. A consequence in
QCD is that non-linear effects occur, resulting in the quark-quark linear potential
that explains the quark confinement, the quarkonia spectra or the Regge trajectories.
Following a parallel between QCD and gravitation, we suggest that these non-linear effects should create an additional logarithmic potential in the classical newtonian
description of gravity. The modified potential may account for the rotation curve of galaxies and other problems, without requiring dark matter.}

\end{abstract}

The strong interaction possesses the particular feature that it confines the
particles that are subject to it, the quarks and the gluons. Quantum Chromodynamic
(QCD) has been established as the quantum field theory for the strong interaction.
Although gravity is in many aspects different from the strong force, we will
exploit the fact that both are non-abelian gauge theories to draw conclusions
on the non-linear effects in gravitation. These are not predicted by the classical
theory of gravitation. Non-linear effects are expected in quantum gravity and
we aim to show that they are also manifest at a macroscopic level. In this paper
we will briefly review some relevant aspects of QCD. We then draw a parallel
between QCD and gravitation and, under this guidance, we predict long distance
effects on gravitation. Using simple heuristics, we establish for a simple case how the gravitation
departs from its 1/r potential due to non-abelian effects {[}\ref{nlinear/nabelian}{]}.
Then, we discuss the implication for dark matter. This communication is not
intended to be a formal derivation of non-abelian effects in gravitation but
rather to trigger more theoretical works on this matter.

QCD can be viewed as the archetype of non-abelian gauge theories: The gluons,
being color charged, interact together. This causes the running coupling constant
\( \alpha _{s} \), to increase with large distances, resulting in
quark confinement and asymptotic freedom which are the two main features of
the strong force. A consequence is that, at distances greater than about one
Fermi, \( \alpha _{s}>1 \). It is then meaningless to use perturbation calculations.
Instead, effective theories (nuclear field theories {[}\ref{walecka}{]}, Regge
phenomenology {[}\ref{regge}{]}, constituent quarks models {[}\ref{isgur_CQM}{]})
or non perturbative techniques (Lattice QCD {[}\ref{walecka}{]}) have to be
used. 

The Regge phenomenology originates from the observation that the hadrons lie
on ``Regge trajectories'', that is to say, the square of the hadron mass \( M \)
is proportional to its spin \( J \): \( J=\alpha _{0}+\alpha 'M^{2} \) where
the ``Regge intercept'' \( \alpha _{0} \), and the slope \( \alpha ' \), characterize
the Regge trajectories. An interpretation of the Regge trajectories was given by Y. Nambu {[}\ref{Nambu}{]}: If the quark-quark potential
is created by a string attaching the two quarks, then the repulsive centrifugal
term, which increases with the spin, makes the baryons of higher spin harder
to bind and thus of higher mass. A string potential would not only explain the
Regge trajectories but also confinement, the energy needed to break the
string being larger than the pion mass, attempts to separate quarks will only result in creating pions. An explanation of the existence of
QCD strings is that the color field lines between the quarks attract themselves
and collapse because of the gluon self-coupling. The collapsed lines add a string-like
term to the quark-quark potential {[}\ref{linepot}{]}:

\begin{equation}
 V_{0}(r)=\frac{\alpha _{s}}{3r}+\sigma r+\iota  
\label{eq:potline}
\end{equation}

\( \sigma  \) is the string tension, \( r \) the distance between the two
quarks and \( \iota  \) a constant. The connection with the Regge trajectory
is \( \alpha '=(2\pi \sigma )^{-1} \). 

The progresses in lattice QCD have shed light on the mechanism of quark confinement
{[}\ref{Bali}{]}, {[}\ref{Lattice}{]}. In particular they have provided a
theoretical ground for the Regge phenomenology, reviving the string picture
in QCD by showing evidences of the existence of flux tubes between quarks. It has also
been shown that the quark-quark potential follows the form of equation (1) and
values of \( \sigma  \) have been computed. The existence of strings and of the potential
(1) are also foreseen by the Dual Meissner Effect {[}\ref{Nambu2}{]}, {[}\ref{Mandelstam}{]},
{[}\ref{t'Hooft}{]} where the QCD vacuum is viewed as a supra-conductor with
the Higgs mechanism responsible for the formation of the strings.  To summarize, the confinement
is due to the non-abelian symmetry of the QCD lagrangian. It results a quark-quark linear potential that can be interpreted in the QCD string picture. 

We turn now our attention to gravity. We make here the usual assumption 
that massless gravitons are the bosons carrying
gravity. In addition, we suppose gravitation to be conceptually close enough to the
 theories describing the three other forces of Nature. We note that this natural assumption is the basis for research on any type of unifying theories such as supersymmetry or superstring theories. The graviton
self-coupling, due to the energy-mass equivalence, makes any theory of quantum
gravity non-abelian. In term of gauge theory, the gravitational field arises
from gauge invariance with respect to displacement transformation {[}\ref{Feyman}{]}
which is non-abelian {[}\ref{deWitt}{]}. We can then expect similar effects
in QCD and gravity. The gravity coupling constant \( G_{N} \)
being very small compared to \( \alpha _{s} \), the non-abelian effects, that
is a term equivalent to \( \sigma r+\iota  \) in equation (1), should become significant
only for very large values of \( r \). Qualitatively, the field lines will get
denser around the straight line linking the two masses. The result will be an
increase of the strength of gravity compared to the newtonian potential in \( 1/r \).
If, from another point of view, we assume the classical potential in \( 1/r \),
then the non-abelian effects will manifest themselves as an apparent increase
of the mass of the object at large distances. The problem of dark matter in
the universe could be simply a manifestation of the non-abelian character
of gravity.

We will now try to estimate the modification to the newtonian potential by using the image of collapsing field lines and doing simple euristics. A non-abelian effect invalidates the classical principle of field superposition
since we have, for example for two fields: Total field=Field\( _{1} \)+Field\( _{2} \)+Interaction
of the fields. We take the case of two masses \( m_{1} \) and \( m_{2} \)
and a newtonian potential. If the field lines are modified because of non-abelian
effects, then the resulting gravitational field at a distance \( r_{i} \) from the
mass \( m_{i} \) can be written as:

\begin{equation}
 \overrightarrow{G}=G_{N}(\frac{m_{1}}{r_{1}^{3}}\overrightarrow{r_{1}}+\frac{m_{2}}{r^{3}_{2}}\overrightarrow{r_{2}})+\overrightarrow{f}_{nA}(m_{1},m_{2},r_{1},r_{2})  
\label{eq:modpot}
\end{equation}

If we could separate \( \overrightarrow{f}_{nA} \) into two independent parts:
\( \overrightarrow{f}_{nA}(m_{1}+m_{2}+r_{1}+r_{2})=\overrightarrow{f}_{1}(m_{1},r_{1})+\overrightarrow{f}_{2}(m_{2},r_{2}) \),
then we could write the field created by one mass as:

\begin{equation}
 \overrightarrow{G_{1}}=\frac{G_{N}m_{1}}{r_{1}^{3}}\overrightarrow{r_{1}}+\overrightarrow{f}_{1}(m_{1},r_{1})  
\label{eq:sepmodpot}
\end{equation}

Apart from the fact that we may not be able to separate \( \overrightarrow{f_{nA}} \),
it may seem incorrect to add a non-abelian term to the field
from one mass since these effects appear only when several fields are present:
For a single mass, the field has a spherical symmetry and the field lines are
not distorted by self-interaction {[}\ref{sphericalsymmetry}{]}. However, we
assume that we can separate \( \overrightarrow{f_{nA}} \) so we can consider
equation (3), the advantage being that it restores the familiar concept of one
potential associated with one source.

We will estimate now \( f_{nA} \). We should have \( f_{nA}\rightarrow 0 \) 
when \( r_{n}\rightarrow 0 \) because close to the mass, the spherical symmetry 
is restored. In order to compute the magnitude of \( f_{1} \), we use the phenomenology
of the collapsed field lines. We set \( m_{1}=m_{2}=M \) and \( 2L \) the
distance between the two masses. We have \( l \) the distance between two field
lines at \( L \), in the case of \( f_{nA}=0 \). If the field lines attract
each other, \( l \) is reduced by \( \Delta l \). \( \Delta l \) should vary
as: 

\begin{equation}
 \Delta l\propto M^{\alpha }G_{N}^{\beta }L^{\gamma }c^{\epsilon }  
\label{eq:deltal}
\end{equation}

with \( c \) the speed of light. The constants \( \alpha  \), \( \beta  \),
\( \gamma  \), \( \varepsilon  \) are to be determined. \( c \) is just here
to regularize the dimensions. We have no prejudice on the signs of \( \alpha  \),
\( \beta  \) and \( \varepsilon  \). However, we have \( \gamma >0 \) since,
as we said, the effects disappear at short distance because of the restored
spherical symmetry. We determine \( \alpha  \), \( \beta  \), \( \gamma  \)
and \( \varepsilon  \) by solving the dimensional equation (4). There is several
solutions, but it appears that if \( \alpha  \), \( \beta  \) are positive
then the only solution is \( \alpha  \)=\( \beta  \)=1, \( \varepsilon =-2 \)
and \( \gamma  \) =0, which is not acceptable since we must have \( \gamma >0 \). If
\( \beta <0 \) then the simplest solution is:

\begin{equation}
 \Delta l\propto \frac{L^{2}c^{2}}{MG_{N}}  
\label{eq:deltal2}
\end{equation}

Now that we have estimated the field line collapse, we can deduce the modification of the potential. The force is proportional to the flux of the field through a surface \( \Delta s \).
We call \( a \) the proportionality constant. The attraction of the field lines
increases the flux through \( \Delta s \). Equivalently, we can keep the flux
fixed and diminish the area \( \Delta s \). If \( \theta  \) is the angle
between the two field lines -without attraction- then
\( \Delta s=\pi [Ltan\theta ]^{2} \) and the diminished area \( \Delta s'=\pi [Ltan(\theta -\Delta \theta )]^{2} \),
where we have \( \Delta \theta =\Delta l/L \). The field magnitude is then:

\begin{equation}
 M_{G}=\frac{1}{a\Delta s'}\simeq \frac{1}{a\pi L^{2}\theta ^{2}}+\frac{2c^{4}}{a\pi G_{N}^{2}M^{2}\theta ^{3}}  
\label{eq:fieldmag}
\end{equation}

The first term is the classical potential. Hence we have \( a\pi \theta ^{2}=1/(G_{N}M) \)
and: 

\begin{equation}
 M_{G}=\frac{G_{N}M}{r^{2}}+\frac{c^{2}\sqrt{a\pi G_{N}M}}{2\sqrt{2}r}  
\label{eq:mg}
\end{equation}

where \( r \) is the distance between the two masses. Equation (6) implies
that the non-abelian effects give rise to a logarithmic potential, less dramatic
than the linear potential in QCD. \( a \) is a constant
to be determined.

To summarize, we have discussed a mechanism that could modify the newtonian law of gravitation,
namely the fact that the field superposition principle is invalidated by
the non-abelian nature of gravitation. These effects being small, the superposition
principle remains valid for most of the cases but for large masses or cosmological distances. 

The modified potential (6) reproduces the observed galaxy rotation curves and solve their
problem without the need of dark matter. To recall the problem, the radial velocity
of the matter in galaxies should follow a keplerian rotation curve, i.e. it
should drop as r\( ^{-1/2} \), as soon as r is large enough, with r the distance
from the galaxy center. However, all the observations of galaxies show a flat
dependence with r or even a positive slope.

In this context, modifications of the newtonian laws have been already proposed
by Milgrom {[}\ref{Milgrom}{]} and then Sanders {[}\ref{Sanders}{]}. These hypothesis are grouped
under the name of MOND, Modified Newtonian Dynamics, see {[}\ref{DMreview2}{]} 
for a recent review. In ref. {[}\ref{Milgrom}{]}
the 2\( ^{nd} \) law of Newton is modified whereas in {[}\ref{Sanders}{]}
it is the gravitational potential. These modifications are empirical. The approaches {[}\ref{Milgrom}{]} and {[}\ref{Sanders}{]} are
equivalent in terms of solving the rotation curve problem, although non-equivalent
in term of physics. Our work supports the ref. {[}\ref{Sanders}{]} approach {[}\ref{badmilgoodsand}{]}.
MOND is characterized by the constant \( a_{0} \). If we replace it by \( (c^{4}\pi a)/8 \),
we find that the equation (6) and the one proposed in {[}\ref{Sanders}{]} are
identical. Taking the most recent determination, \( a_{0}=1.2\times 10^{-10} \)
m.s\( ^{-2} \) {[}\ref{a0}{]}, we obtain \( a=4\times 10^{-44} \) m\( ^{-3} \)s\( ^{2} \).

Several authors have systematically studied the MOND hypothesis. We summarize
here their work since they provide the experimental evidence supporting
equation (6). Some other supportive comparisons with observations are discussed in {[}\ref{DMreview2}{]}.

A) Rotation curves of the different varieties of galaxies are fit with the
parameter \( a_{0} \), adjusted once for all {[}\ref{DMreview}{]}.

B) The prediction by Milgrom of the existence of positive slopes for the rotation
curves was experimentally confirmed {[}\ref{positiveslopes}{]}.

C) The origin of the Tully-Fisher relation {[}\ref{tullyfisher}{]}, linking the
luminosity of a galaxy to its optical rotational velocity at the power 3 or
4, is disputed {[}\ref{tullyfisher_review}{]}.
However, this relation is a straightforward consequence
of a gravity obeying the equation (6) {[}\ref{DMreview}{]}.

D) At larger scale, MOND explains the data on binary system of galaxies, small
cluster and large clusters without need of introducing dark matter {[}\ref{Milgrom2}{]}.

MOND appears as a solid alternative, or at least a complement, to dark matter. However, it suffers 
from is lack of theoretical justification. Our work provides a possible 
theoretical basis for such model. However, a major difference between MOND and our work is that MOND modifies the gravitation potential in one way while in our case the modification depends upon the distribution of mass. Hence we expect MOND to fail describing data involving particular distribution of matter while calculations following our work might be able to explain them.  

Before to summarize, we should point out that we used the QCD string picture as a guidance but the rightness of the reasonning is not contigent to the existance of QCD strings. Rather the important point is that non-abelian symmetry is the origin of confinement, which is an established fact. It is known, for example, that the peculiar running of \( \alpha _{s} \) is due to the gluons self-coupling that create the anti-screening causing \( \alpha _{s} \) to grow with the distance.    

To summarize, following a parallel between QCD and gravitation, namely the fact
that both theories are non-abelian, we have suggested that non-abelian effects
can be described by an additional logarithmic term in the newtonian potential.
As shown in the context of the purely empirical MOND model, the 
modified resulting potential explains cosmological problems otherwise 
requiring a large amount of dark matter. In the same context, it has also 
been pointed out that eq. (6) does not contradict general relativity nor it precludes the existence of a small amount of dark matter.


\begin{references}
\bibitem{17}\label{sphericalsymmetry}Without using the spherical symmetry argument: the
interactions of two lines sandwiching a third line will compensate, yielding
a total null effect on this third line.
\bibitem{19}\label{badmilgoodsand} In addition it is been argued in {[}\ref{DMreview}{]}
that the modification of the gravitational potential is preferable over a modification
of the 2\( ^{nd} \) law, which would modify all the newtonian dynamics.
\end{references}
\end{document}